\documentclass[aip,amsmath,amssymb,graphicx]{revtex4-1}

\usepackage{graphicx}
\usepackage{dcolumn}
\usepackage{bm}

\usepackage[utf8]{inputenc}
\usepackage[T1]{fontenc}
\usepackage{mathptmx}
\usepackage{etoolbox}

\usepackage{xcolor} 
\usepackage{soul}   

\draft 
\begin{document}

\preprint{}

\title{Triggering of extreme events and coherent-structure modulation in wall-turbulence under cyclostationary forces}

\author{Ao Xu}
\affiliation{ 
Institute of Extreme Mechanics, School of Aeronautics, Northwestern Polytechnical University, Xi'an 710072, China
}%
\affiliation{ 
National Key Laboratory of Aircraft Configuration Design, Key Laboratory for Extreme Mechanics of Aircraft of Ministry of Industry and Information Technology, Xi’an 710072, China
}%
\author{Yun-Qian Bi}%
\affiliation{ 
Institute of Extreme Mechanics, School of Aeronautics, Northwestern Polytechnical University, Xi'an 710072, China
}%

\author{Heng-Dong Xi}
 \email[Author to whom correspondence should be addressed:]{hengdongxi@nwpu.edu.cn}
\affiliation{ 
Institute of Extreme Mechanics, School of Aeronautics, Northwestern Polytechnical University, Xi'an 710072, China
}%
\affiliation{ 
National Key Laboratory of Aircraft Configuration Design, Key Laboratory for Extreme Mechanics of Aircraft of Ministry of Industry and Information Technology, Xi’an 710072, China
}%

\date{\today}

\begin{abstract}
Atmospheric gusts expose wall-bounded turbulence to severe unsteady forcing, triggering complex non-equilibrium dynamics and extreme aerodynamic loads.
In this study, direct numerical simulations are performed to investigate the spatiotemporal modulation of turbulent structures and the triggering mechanisms of near-wall extreme events under Gaussian-type transient forcing.
The results reveal that high-amplitude gusts inject energy primarily into the streamwise velocity component, inducing a pronounced non-equilibrium phase lag during turbulent energy redistribution. This process produces hysteresis in wall friction and extends the relaxation time.
Spectral and continuous wavelet analyses demonstrate that intense gust forcing suppresses high-frequency random fluctuations and reorganizes turbulent kinetic energy into low-frequency coherent structures. The characteristic frequency of these energetic structures locks onto the gust driving frequency, with a relative deviation of only $2.4\%$.
Furthermore, the occurrence probability of extreme near-wall events, including extreme positive (EP) wall-shear-stress events and rare backflow (BF) events, increases by up to an order of magnitude under severe forcing.
Using a two-step conditional averaging technique, we demonstrate that BF events are actively driven by intense, localized adverse pressure gradients and energetic ejections, which promote spanwise vortex roll-up in the buffer layer.
By contrast, EP events are governed by energetic sweeps of high-speed fluid that compress intense spanwise vorticity into the immediate vicinity of the wall.
These findings provide physical insights into non-equilibrium energy transfer and offer theoretical guidance for load alleviation and robust flow control of unmanned aerial vehicles operating in unsteady atmospheric environments.
\footnote{
This article may be downloaded for personal use only.
Any other use requires prior permission of the author and AIP Publishing.
This article appeared in Xu \emph{et al.}, Phys. Fluids \textbf{38}, 065149 (2026) and may be found at \url{https://doi.org/10.1063/5.0337822}.
}
\end{abstract}


\maketitle 

\section{Introduction}
Unsteady flows in the atmospheric boundary layer (ABL) are characterized
by multiscale vortical structures and intermittent fluctuations
that pose challenges for various engineering applications \cite{mahrt2014stably,stevens2017flow}.
Among atmospheric disturbances, gusts, which are defined as sudden, high-amplitude increases
in wind speed followed by rapid decay, represent a particularly severe threat \cite{jones2022physics}.
These transient events can degrade the aerodynamic performance,
structural integrity, and control stability of aircraft and unmanned aerial vehicles
(UAVs) operating in the lower atmosphere \cite{watkins2006atmospheric,mohamed2014fixed}.
Unlike continuous harmonic disturbances,
isolated gusts impose impulsive external forcing on turbulent boundary layers.
During a gust encounter,
traditional quasi-steady aerodynamic assumptions frequently break down.
The resulting rapid variations in the magnitude and direction of the incoming flow
induce nonlinear, unsteady aerodynamic loads \cite{leishman2002challenges,corke2015dynamic}.
Furthermore, such abrupt forcing can trigger flow separation, dynamic stall,
and extreme wall-shear-stress fluctuations, thereby exacerbating structural fatigue
and compromising flight safety \cite{granlund2014airfoil,andreu2022mitigation}.
Consequently, transient atmospheric disturbances disrupt
the equilibrium state of wall-bounded turbulence
and drive the flow into a non-equilibrium state.
It is therefore essential to elucidate the spatiotemporal evolution of turbulent coherent structures
and the associated energy-transfer mechanisms under transient forcing.
Such an understanding is a prerequisite for developing accurate predictive models \cite{xu2026revisit}
and advanced flow-control strategies.

Canonical wall-bounded turbulence subjected to time-varying external forcing
provides an idealized framework for isolating the physics underlying gust--flow interactions.
In statistically steady wall-bounded flows, near-wall dynamics are characterized by low-speed streaks and quasi-streamwise vortices \cite{kim1987turbulence,jeong1997coherent,massaro2026interpreting}. 
These coherent structures undergo a self-sustaining process (SSP) involving continuous generation, breakdown, and regeneration \cite{hamilton1995regeneration,waleffe1997self,jimenez1999autonomous}. 
This structural organization extends from near-wall streak instabilities \cite{schoppa2002coherent} to larger-scale hairpin vortex packets \cite{adrian2007hairpin} and self-sustained motions in the outer layer \cite{hwang2010self,smits2013wall}. 
These canonical structures govern turbulent kinetic energy (TKE) production and wall-normal momentum transport through distinct bursting events. 
The roles of ejections and sweeps in this process were first identified through flow-visualization studies \cite{kline1967structure,corino1969visual} and later through structural classifications \cite{robinson1991coherent}. 
More recently, small-scale near-wall bursting events have been shown to be strongly modulated by large-scale structures \cite{mathis2009large}.
Under severe unsteady external forcing, however,
the classical equilibrium of the turbulent energy cascade can be disrupted,
giving rise to complex, phase-dependent modulation across multiple spatiotemporal scales.

Extensive research on non-stationary wall-bounded turbulence has primarily focused on periodic pulsating flows and transient step changes. 
Early studies of transient pipe flows and accelerating boundary layers established a foundational understanding of unsteady turbulent modulation \cite{he2000study,piomelli2000turbulent}. 
Subsequent numerical simulations of pulsating channel flows showed that the structural response of turbulence depends strongly on the frequency and amplitude of the imposed oscillation \cite{scotti2001numerical,manna2015pulsating,ebadi2019mean}, and that appropriately tuned transient forcing can even lead to sustained turbulence suppression \cite{scarselli2023turbulence}. 
A phase lag between the external forcing and the near-wall turbulent response is commonly observed. 
This hysteresis is generally attributed to the finite relaxation time required for turbulent kinetic energy (TKE) to be redistributed from energetic large scales in the outer region to dissipative scales near the wall \cite{lozano2012three}. 
Despite these insights, realistic atmospheric gusts differ from continuous harmonic pulsations \cite{jones2022physics}. 
Whereas harmonic forcing produces continuously recurring cyclostationary cycles, isolated atmospheric gusts impose high-amplitude transient perturbations that drive the boundary layer away from equilibrium. 
This process induces a pronounced memory effect during both the acceleration and relaxation phases \cite{tardu1993wall,he2000study,greenblatt2004rapid}. 
Consequently, the transient accumulation, spectral reorganization, and rapid release of turbulent energy induced by isolated gusts remain insufficiently understood.

A critical manifestation of this non-equilibrium energy transfer under strong modulation is the occurrence of near-wall extreme events. 
Recent high-fidelity direct numerical simulation (DNS) studies have systematically characterized extreme wall-shear-stress events \cite{hutchins2011three,smits2013wall,blonigan2019extreme}. 
Extreme positive (EP) wall-shear-stress events are localized phenomena associated with intense sweep (Q4) motions, in which high-speed fluid impinges on the wall \cite{sheng2009buffer,pan2018extremely}. 
Conversely, backflow (BF) events, characterized by negative streamwise wall-shear stress, represent localized flow reversals. 
Although extreme events are rare in canonical statistically steady turbulence (e.g., occurring with a probability of less than $0.1\%$ at $Re_{\tau} \approx 1000$), they serve as key indicators of flow intermittency \cite{guerrero2020extreme,orlu2011fluctuating,vinuesa2017characterisation}. 
These rare events are closely associated with intense, localized adverse pressure gradients that strongly modulate near-wall turbulent structures \cite{krogstad1995influence,bross2019interaction}. 
The three-dimensional topology and kinematic features of such events have been well documented in canonical channels and pipes \cite{lenaers2012rare,cardesa2019structure,chin2018flow}, as well as during boundary layer transition \cite{wu2020negative}. 
More recently, advanced volumetric conditional-averaging techniques have extended the understanding of extreme wall-shear-stress and wall-pressure-fluctuation events to complex environments, including supersonic and hypersonic boundary layers \cite{wan2023conditional}. 
Despite these insights from statistically steady regimes, the manner in which a high-amplitude, time-varying gust modulates the occurrence frequency, spatial extent, and vortical triggering mechanisms of these extreme events remains an open question.

To bridge this gap, we perform DNS to investigate the structural evolution and triggering mechanisms of extreme events in a turbulent shear flow driven by an unsteady, Gaussian-type external force. 
This Gaussian forcing model is designed to emulate the rapid acceleration and subsequent relaxation characteristic of severe atmospheric gusts. 
By independently varying the forcing amplitude and duration, we aim to uncover the phase-dependent modulation mechanisms of the turbulent flow. 
It is also important to clarify the modeling rationale adopted in this study. 
In realistic atmospheric environments, gusts appear as complex disturbances characterized by spatial heterogeneity and variable convective velocities. 
Recent experimental studies \cite{du2024characteristics} have shown that generating prescribed gust profiles in laboratory settings can introduce spatial distortions and velocity nonuniformity, primarily owing to wakes generated by gust-producing devices. 
Furthermore, capturing and predicting the nonlinear spatiotemporal evolution of such spatially heterogeneous gusts often requires advanced data-driven techniques, such as hybrid autoencoder--random-forest algorithms \cite{du2025data}. 
Consequently, resolving these spatial features may obscure the underlying flow physics by making it difficult to distinguish turbulent-structure modulation driven by spatial convection from that driven by temporal unsteadiness. 
To address this issue, the present study employs a temporal surrogate model in which the gust is idealized as a spatially uniform, time-varying body force. 
This canonical abstraction decouples temporal unsteadiness from spatial convective effects, allowing us to isolate the non-equilibrium structural reorganization mechanisms without the confounding influence of spatial wake interference.

The primary objectives of this work are threefold: (i) to elucidate the spatiotemporal evolution of coherent structures and cross-scale energy transfer during a gust encounter; (ii) to clarify the phase-averaged modulation mechanisms and memory effects in turbulent statistics under transient forcing; and (iii) to identify and characterize the three-dimensional topological origins of gust-triggered extreme events (BF and EP). 
The remainder of the paper is organized as follows. 
Section ~\ref{sec2} describes the mathematical model, the DNS setup, and the formulation of the Gaussian forcing. 
Section ~\ref{sec3} presents the macroscopic statistical response, multiscale spectral modulation, and topological characterization of near-wall extreme events. 
Finally, Sec.~\ref{sec4} summarizes the main conclusions.

\section{\label{sec2}Numerical methods}
\subsection{Governing equations and gust-forcing model}
Direct numerical simulations (DNS) are conducted for a fully developed turbulent channel flow subjected to time-varying, Gaussian-type external forcing.
The fluid is assumed to be Newtonian and incompressible.
The governing equations are given by:
\begin{align}
    \nabla \cdot \mathbf{u} &= 0, \label{eq:continuity} \\
    \frac{\partial \mathbf{u}}{\partial t} + \mathbf{u} \cdot \nabla \mathbf{u} &= -\frac{1}{\rho}\nabla p + \nu \nabla^2 \mathbf{u} + F_{gust}(t)\hat{\mathbf{x}}, \label{eq:momentum}
\end{align}
where $\mathbf{u}=(u,v,w)$ denotes the velocity vector, with components in the streamwise $(x)$, wall-normal $(y)$, and spanwise $(z)$ directions, respectively.
The quantities $\rho$, $p$, and $\nu$ denote the fluid density, pressure, and kinematic viscosity, respectively.
The term $F_{gust}(t)$ denotes the unsteady external forcing applied in the streamwise direction $\hat{\mathbf{x}}$.
Numerically, this spatially uniform transient body force acts as an equivalent unsteady pressure gradient.
The application of a time-varying body force to fully developed channel flow is a well-established framework for investigating non-stationary wall-bounded flows \cite{he2000study,greenblatt2004rapid,manna2012pulsating}.
Although this formulation shares similarities with pulsating channel flow, the present study employs a Gaussian-type pulse rather than a continuous harmonic oscillation.
This choice mimics the rapid, isolated intensification and subsequent relaxation characteristic of atmospheric gusts, thereby driving wall-bounded turbulence into a non-equilibrium state.

\begin{figure}
\includegraphics[width=0.9\textwidth]{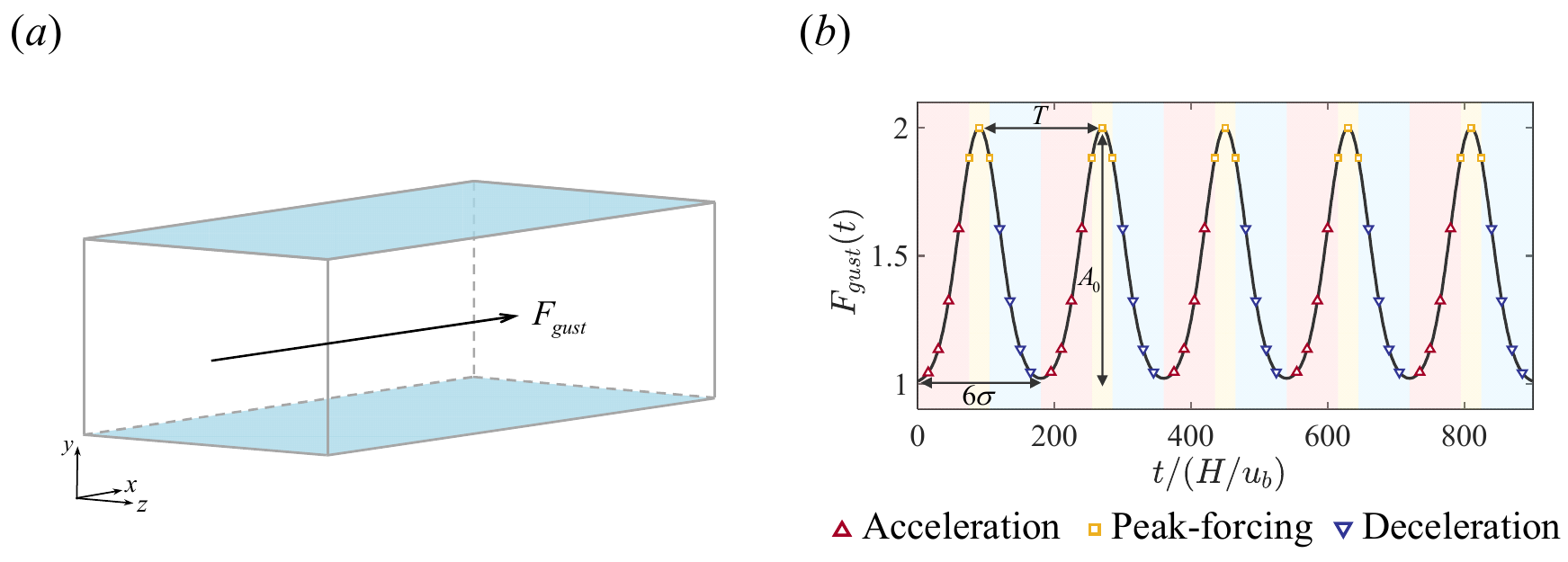}
\caption{\label{fig1} Schematic of the computational domain and the time-varying gust-forcing model. 
($a$) Geometry of the turbulent channel flow with a spatially homogeneous streamwise body force $F_{gust}$. 
($b$) Time history of the periodic Gaussian-type gust forcing $F_{gust}(t)$, 
partitioned into acceleration (red), peak-forcing (yellow), and deceleration (blue) phases.}
\end{figure}

The atmospheric gust is modeled as a transient acceleration followed by a relaxation phase, as illustrated in Fig.~\ref{fig1}.
To capture this rapid intensification and subsequent decay, $F_{gust}(t)$ is prescribed as a Gaussian pulse,
\begin{equation}
F_{gust}(t)=\left\{1+\sum_{i=1}^{N} A_{0,i} \exp\left[-\frac{(t-t_{0,i})^2}{2\sigma_i^2}\right]\right\} F_{steady},
\end{equation}
where $F_{steady}$ is the constant equivalent mean pressure gradient required to maintain the baseline statistically steady turbulent channel flow at the nominal friction Reynolds number $Re_{\tau}$.
For the $i$th Gaussian pulse ($i=1,\ldots,N$), $A_{0,i}$ denotes the peak amplitude, $t_{0,i}$ denotes the temporal center (i.e., the instant of maximum intensity), and $\sigma_i$ determines the duration, or characteristic timescale, through the pulse width.
Although atmospheric gusts encountered by aircraft are inherently isolated transient events, simulating only a single isolated gust in DNS would provide insufficient samples for extracting statistically converged coherent structures.
To circumvent this limitation, we employ a sequence of $N$ periodic Gaussian pulses with identical amplitudes ($A_{0,i}=A_0$) and durations ($\sigma_i=\sigma$) [as illustrated in Fig.~\ref{fig1}(b)], thereby rendering the flow cyclostationary.
In this cyclostationary setup, successive pulses are arranged periodically such that $t_{0,i+1}-t_{0,i}=T$, where $T$ is the forcing period.
In cyclostationary turbulence, the macroscopic statistical moments vary with time but repeat periodically, satisfying $\langle q(\mathbf{x},t)\rangle=\langle q(\mathbf{x},t+T)\rangle$, where $q$ represents an arbitrary flow quantity and $\langle \cdot \rangle$ denotes phase averaging.

For the phase-resolved analysis, the temporal evolution of a single gust event is defined based on the properties of the Gaussian distribution.
According to the empirical three-sigma rule, the interval within $\pm 3\sigma$ of the mean contains more than $99.7\%$ of the total pulse contribution.
Outside this interval, the gust perturbation decays to a negligible magnitude and exerts no significant influence on the background turbulence.
Accordingly, the effective duration of a single gust event is defined as $6\sigma$, spanning the temporal interval $t\in[t_0-3\sigma,t_0+3\sigma]$.
This temporal window is sufficiently broad to ensure the dynamic independence of consecutive gust cycles.
Although the flow exhibits a memory effect associated with turbulent relaxation, the $6\sigma$ cycle window provides sufficient time for the non-equilibrium turbulent kinetic energy to dissipate.
Consequently, the boundary layer relaxes to the canonical statistically steady baseline state before the onset of the subsequent pulse, as corroborated later by the recovery of the macroscopic skin friction to its steady-state value at the cycle boundaries.
Thus, although a cyclostationary sequence is employed to obtain statistically converged samples, each forcing cycle behaves as an isolated transient gust encounter.
To facilitate a unified comparison of turbulent responses across different timescales (i.e., different $\sigma$ values), we introduce a dimensionless phase parameter, $\Phi=(t-t_0+3\sigma)/(6\sigma)$.
This definition maps the effective gust cycle onto the normalized interval $\Phi\in[0,1]$.
Based on the temporal derivative of the external driving force, $\mathrm{d}F_{gust}(t)/\mathrm{d}t$, the gust cycle is further partitioned into three dynamic stages:
(i) the acceleration phase ($t\in[t_0-3\sigma,t_0-0.5\sigma]$, corresponding to $\Phi\in[0,0.42]$), during which the temporal derivative is positive; 
(ii) the peak-forcing phase ($t\in[t_0-0.5\sigma,t_0+0.5\sigma]$, corresponding to $\Phi\in[0.42,0.58]$), which encompasses the vicinity of maximum forcing, where the temporal derivative changes sign; and 
(iii) the deceleration phase ($t\in[t_0+0.5\sigma,t_0+3\sigma]$, corresponding to $\Phi\in[0.58,1]$), during which the temporal derivative is negative.
To investigate the transient modulation of turbulent structures, $A_0$ and $\sigma$ are varied independently to simulate gusts with different strengths and timescales.

\subsection{Numerical discretization and computational domain}
The governing equations are solved using an in-house multiple-relaxation-time lattice Boltzmann method (MRT-LBM) code.
The lattice Boltzmann method is adopted as the numerical framework for the DNS of turbulent flows.
To recover the macroscopic Navier--Stokes equations, the density distribution function evolves according to
\begin{equation}
f_i(\mathbf{x}+\mathbf{e}_i\delta_t,t+\delta_t)-f_i(\mathbf{x},t)=-(\mathbf{M}^{-1}\mathbf{S})_{ij}\left[\mathbf{m}_j(\mathbf{x},t)-\mathbf{m}_j^{(\mathrm{eq})}(\mathbf{x},t)\right]+\delta_t F_i^{\prime},
\end{equation}
where $f_i$ denotes the density distribution function, $\mathbf{x}$ is the spatial coordinate, $t$ is time, and $\delta_t$ is the time step.
The vector $\mathbf{e}_i$ denotes the discrete velocity along the $i$th lattice direction.
The matrix $\mathbf{M}$ is a $19 \times 19$ orthogonal transformation matrix associated with the D3Q19 discrete-velocity model.
The equilibrium moments $\mathbf{m}^{(\mathrm{eq})}$ are given by
\begin{equation}
\begin{split}
    \mathbf{m}_{\mathrm{D3Q19}}^{(\mathrm{eq})} = \rho \bigg[ & 1, -11 + 19|\mathbf{u}|^2, 3 - \frac{11}{2}|\mathbf{u}|^2, u, -\frac{2}{3}u, v, \\
    & -\frac{2}{3}v, w, -\frac{2}{3}w, 2u^2 - v^2 - w^2, -\frac{1}{2}(2u^2 - v^2 - w^2), v^2 - w^2, \\
    & -\frac{1}{2}(v^2 - w^2), uv, vw, uw, 0, 0, 0 \bigg]^T.
\end{split}
\label{eq:m_eq_D3Q19}
\end{equation}
In the present MRT-LBM solver, Guo's forcing scheme \cite{guo2002discrete} is adopted. 
This scheme has been shown to eliminate low-order errors caused by the discretization of external forces and to exhibit second-order accuracy in both time and space.
The kinematic viscosity in lattice units is given by $\nu_f=(s_\nu^{-1}-0.5)/3$.
The macroscopic fluid density $\rho_f$ and velocity $\mathbf{u}_f$ are recovered as
$\rho_f=\sum_{i=0}^{18}f_i$ and $\mathbf{u}_f=(\sum_{i=0}^{18}\mathbf{e}_i f_i+\mathbf{F}/2)/\rho_f$, respectively.
Further details of the LB method and the validation of the in-house code can be found in our previous work \cite{xu2017accelerated,xu2023multi}.

The simulations are performed in a computational domain of size $L_x \times L_y \times L_z = 4h \times 2h \times 2h$, where $h=H/2$ denotes the channel half-height, and $H$ is the channel height.
Periodic boundary conditions are imposed in the streamwise and spanwise directions, whereas no-slip boundary conditions are applied at the upper and lower walls.
The baseline statistically steady flow is maintained at a nominal friction Reynolds number of $Re_{\tau}=u_{\tau} h/\nu =180$, where $u_{\tau}$ is the friction velocity of the unperturbed flow. 
Under these reference conditions, the domain dimensions correspond to $L_x^+ \approx 720$ and $L_z^+ \approx 720$ in wall units.
This domain is substantially larger than the theoretical minimal-flow-unit requirements ($L_x^+\approx300$, $L_z^+\approx100$) and is therefore sufficient to sustain canonical near-wall coherent structures without spurious periodic interference, as confirmed by the statistical validation presented in Sec.~\ref{sec2.3}.
The computational domain is uniformly discretized using a Cartesian lattice grid of $N_x \times N_y \times N_z = 401 \times 211 \times 201$.
A fully developed turbulent field at $Re_\tau=180$, precomputed over $150$ eddy turnover times ($H/u_\tau$), corresponding to three million LBM iteration steps, serves as the initial condition for all subsequent gust simulations.
Throughout the cyclostationary simulations, a globally constant time step and a uniform grid spacing are employed to maintain standard Cartesian lattice alignment.
To ensure the statistical significance of the results, particularly for capturing rare extreme events, the simulations are conducted over extended physical durations. 
Depending on the gust duration $\sigma$, 25 to 30 independent gust cycles are simulated, yielding a total sampling time of up to $4500 H/u_b$. 
Both long-time-averaged and phase-averaged statistics are verified to be converged.

To ensure the fidelity of the DNS, the grid resolution must be assessed not only for the baseline flow but also at the instantaneous peak of the most severe gust.
Under the strongest forcing condition ($A_0=5.0$, $\sigma=20$), the flow undergoes intense acceleration, causing the instantaneous friction Reynolds number, $Re_{\tau}^{\mathrm{peak}}$, to reach approximately $388.8$.
Consequently, the domain size expressed in inner units increases (e.g., $L_x^+ \approx 1555$ at the peak), thereby providing sufficient spatial extent to accommodate elongated transient structures.

To quantitatively verify the grid resolution under varying gust severities, Table~\ref{tab:grid_resolution} compares the resolution metrics at the steady-state and peak-$Re_{\tau}$ conditions. 
We conditionally average the instants at which $Re_\tau$ reaches its maximum within each gust cycle to obtain $\langle Re_\tau\rangle_{peak}$. 
Based on the local friction velocity $u_\tau$, the near-wall grid spacing $\langle\Delta y_{min}^+\rangle_{peak}$ is calculated; based on the local Kolmogorov scale $\eta_K$, the maximum ratio of the global grid spacing to the dissipation scale, $\langle\Delta y_{max}/\eta_K\rangle_{peak}$, is determined. 
Similarly, all instants during the steady phase of the gust driving force are conditionally averaged to obtain $\langle\Delta y_{min}^+\rangle_{steady}$ and $\langle\Delta y_{max}/\eta_K\rangle_{steady}$.

\begin{table}[htbp]
\caption{\label{tab:grid_resolution} Comparison of grid-resolution metrics at the steady-state and peak-$Re_{\tau}$ conditions under different gust parameters.}
\begin{ruledtabular}
\begin{tabular}{ccccccc}
$A_0$ & $\sigma$ & $\langle Re_\tau\rangle_{peak}$ & $\langle\Delta y_{min}^+\rangle_{peak}$ & $\langle\Delta y_{max}/\eta_K\rangle_{peak}$ & $\langle\Delta y_{min}^+\rangle_{steady}$ & $\langle\Delta y_{max}/\eta_K\rangle_{steady}$ \\
\hline
0.2 & 10 & 203.8 & 0.97 & 1.25 & 0.87 & 1.13 \\
0.2 & 20 & 201.3 & 0.95 & 1.20 & 0.84 & 1.07 \\
0.2 & 30 & 206.3 & 0.98 & 1.31 & 0.84 & 1.12 \\
0.5 & 10 & 218.9 & 1.04 & 1.36 & 0.94 & 1.25 \\
0.5 & 20 & 216.0 & 1.03 & 1.37 & 0.95 & 1.16 \\
0.5 & 30 & 220.7 & 1.05 & 1.40 & 0.93 & 1.25 \\
1.0 & 10 & 238.3 & 1.13 & 1.48 & 0.97 & 1.40 \\
1.0 & 20 & 239.2 & 1.13 & 1.49 & 1.03 & 1.29 \\
1.0 & 30 & 245.1 & 1.16 & 1.49 & 1.04 & 1.30 \\
5.0 & 20 & 388.8 & 1.90 & 2.50 & 1.54 & 2.03 \\
\end{tabular}
\end{ruledtabular}
\end{table}

As shown in the last row of Table~\ref{tab:grid_resolution}, under the highest gust amplitude ($\sigma=20$, $A_0=5.0$), for which the transient peak reaches $\langle Re_\tau\rangle_{peak} \approx 388.8$, the maximum ratio of the grid spacing to the local Kolmogorov scale, $\langle\Delta y_{max}/\eta_K\rangle_{peak}$, is 2.50. 
Classical turbulence theory indicates that turbulent kinetic energy dissipation is complete when the dimensionless wavenumber satisfies $\kappa\eta_K>1$ \cite{pope2001turbulent}. 
According to the relation between wavenumber and spatial scale, the corresponding wavelength is $\lambda=2\pi/\kappa<2\pi\eta_K\approx6.28\eta_K$. 
Based on the Nyquist criterion, the critical resolvable scale is $\Delta x=\lambda/2<3.14\eta_K$. 
Therefore, the energy contained at scales smaller than $3\eta_K$ is negligible. 
Consequently, a maximum grid spacing of $2.50\eta_K$ is sufficient to resolve the spectral range that dominates turbulent energy dissipation. 
Furthermore, the LBM exhibits low numerical dissipation and dispersion. 
Thus, the grid resolution employed in this study accurately resolves the small-scale turbulent structures in the dissipation range, even under the most severe unsteady forcing.

\subsection{\label{sec2.3}Validation of the baseline flow}
\begin{figure}
\includegraphics[width=0.9\textwidth]{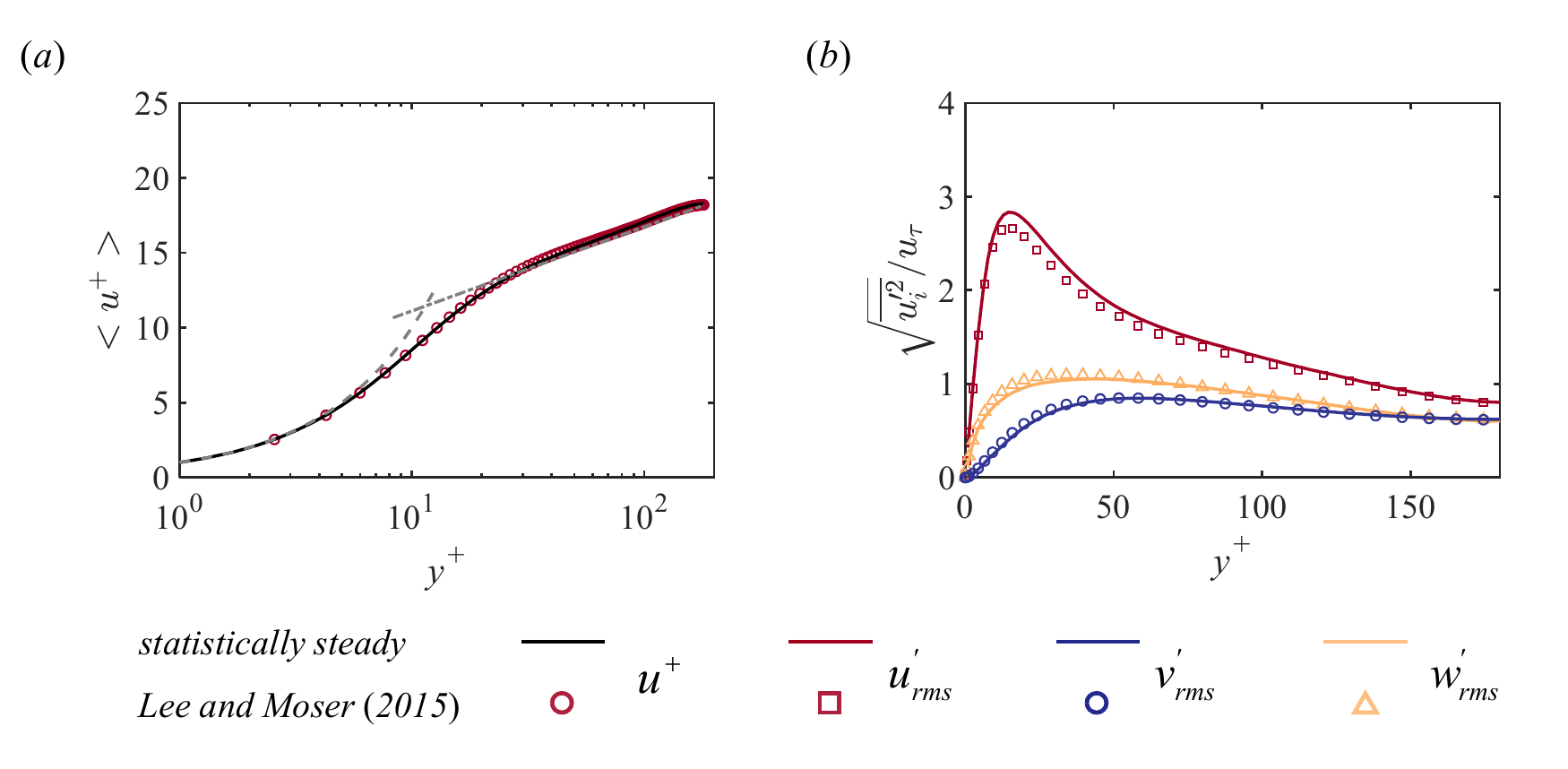}
\caption{\label{fig2} Validation of the baseline DNS setup at $Re_\tau \approx 180$ against the benchmark data of Lee and Moser \cite{lee2015direct}. 
($a$) Mean streamwise velocity profile, $\langle u^+\rangle$. ($b$) Root-mean-square (RMS) velocity fluctuations. Solid lines denote the present results; open symbols denote the benchmark data.}
\end{figure}

To verify the fidelity of the numerical setup, the baseline statistically steady turbulent channel flow ($A_0=0$) is validated against the benchmark DNS data of \citet{lee2015direct}. 
As shown in Fig.~\ref{fig2}, both the mean streamwise velocity profile ($\langle u^+\rangle$) and the root-mean-square (RMS) velocity fluctuations ($u_{rms}^{\prime}$, $v_{rms}^{\prime}$, and $w_{rms}^{\prime}$) agree well with the reference spectral DNS data throughout the channel.
This validation confirms that the grid resolution, numerical scheme, and domain size are sufficient for accurately capturing the baseline turbulent statistics. 
Consequently, this steady-state flow provides a reliable foundation for evaluating the structural modulation induced by transient gust forcing in Secs. \ref{sec3.1} and \ref{sec3.2}.

\section{\label{sec3}Results and discussion}
\subsection{\label{sec3.1}Response of wall-bounded turbulence to gust forcing}
\begin{figure}
\includegraphics[width=\textwidth]{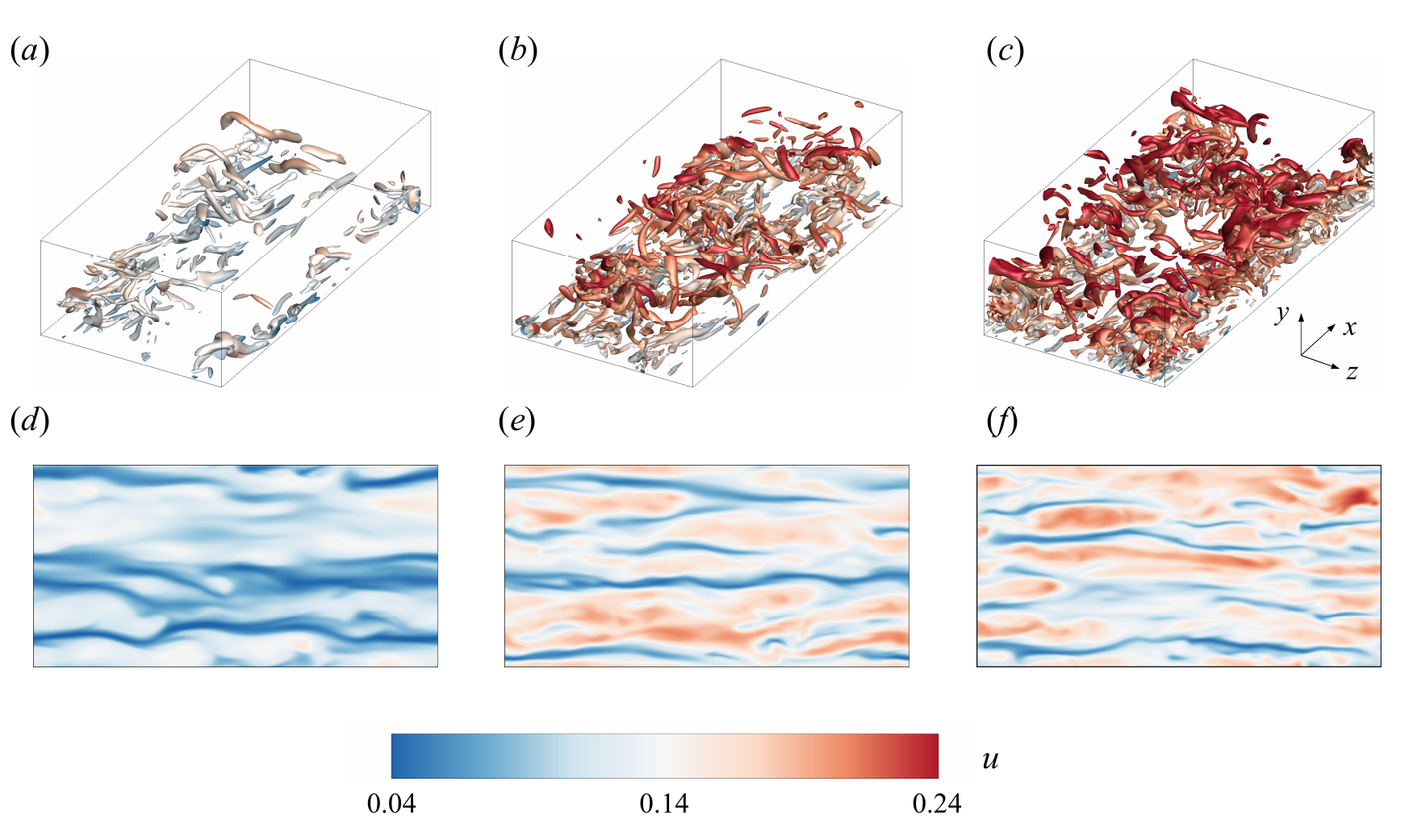}
\caption{\label{fig3} Instantaneous coherent structures and near-wall velocity streaks under extreme gust forcing ($\sigma=20$, $A_0=5.0$). 
($a$)--($c$) Vortical structures identified by the $Q$-criterion ($Q^\ast=0.1$) and colored by the streamwise velocity $u$. 
($d$)--($f$) Streamwise velocity contours in a near-wall $x$--$z$ plane located at $y^+ \approx 15$. The panels correspond to the acceleration ($a$) and ($d$), peak-forcing ($b$) and ($e$), and deceleration ($c$) and ($f$) phases.}
\end{figure}

To visualize the structural evolution of the flow, instantaneous snapshots of vortical structures identified by a positive isovalue of the \emph{Q}-criterion are presented in Figs.~\ref{fig3}($a$)--($c$).
The topology of these coherent structures exhibits strong phase dependence.
During the acceleration phase [see Fig.~\ref{fig3}($a$)], the vortical structures are relatively sparse and disorganized.
At this stage, the momentum introduced by the gust is still accumulating in the bulk flow, and the near-wall shear layer has not yet developed strong instabilities.
As the gust approaches its peak and enters the early deceleration phase [see Fig.~\ref{fig3}($b$)], the flow reorganizes, and the density of hairpin-like vortices and quasi-streamwise vortex tubes increases markedly, resulting in dense vortical clusters.
This stage corresponds to the peak of turbulent kinetic energy (TKE) production, during which the accumulated momentum is converted into intense turbulent fluctuations.
Notably, during the deceleration phase [see Fig.~\ref{fig3}($c$)], the flow retains a high degree of spatial complexity and vortical density.
This sustained structural intensity illustrates the memory effect in non-stationary turbulence, whereby the dissipation of coherent structures lags behind the decay of the external forcing.

To quantitatively understand this memory effect, the gust timescale is compared with the reorganization timescale of the near-wall coherent structures. 
A complete cycle of the canonical self-sustaining process (SSP) \cite{jimeneza2005characterization} requires a characteristic period of $T_{SSP}^+ \approx 400$. 
Across the varying gust conditions prescribed in this study, the effective acceleration period spans $T_{accel}^+ \approx 470$, $526$, and $618$ for amplitudes $A_0=0.2$, $0.5$, and $1.0$, respectively. 
Because this gust acceleration phase is of the same order of magnitude as a single SSP cycle $T_{accel}^+ \sim O(T_{SSP}^+)$, the turbulent structures lack sufficient time to undergo the multiple evolutionary iterations required to redistribute the newly injected momentum and reach a new equilibrium. 
Consequently, turbulent momentum exchange substantially lags the rapid external momentum input, generating pronounced hysteresis that becomes stronger at larger amplitudes $A_0$. 
The corresponding instantaneous streamwise velocity contours at $y^+ \approx 15$ [see Figs.~\ref{fig3}($d$)--($f$)], which serve as near-wall footprints of the phase lag, visually corroborate this modulation.
While the near-wall streaks appear elongated and predominantly low-speed during the acceleration phase, they become increasingly sinuous during the deceleration phase. 
This behavior indicates enhanced streak instability and subsequent bursting events, which serve as the primary precursors of the extreme events analyzed in Sec. \ref{sec3.2}.

\begin{figure}
\includegraphics[width=0.75\textwidth]{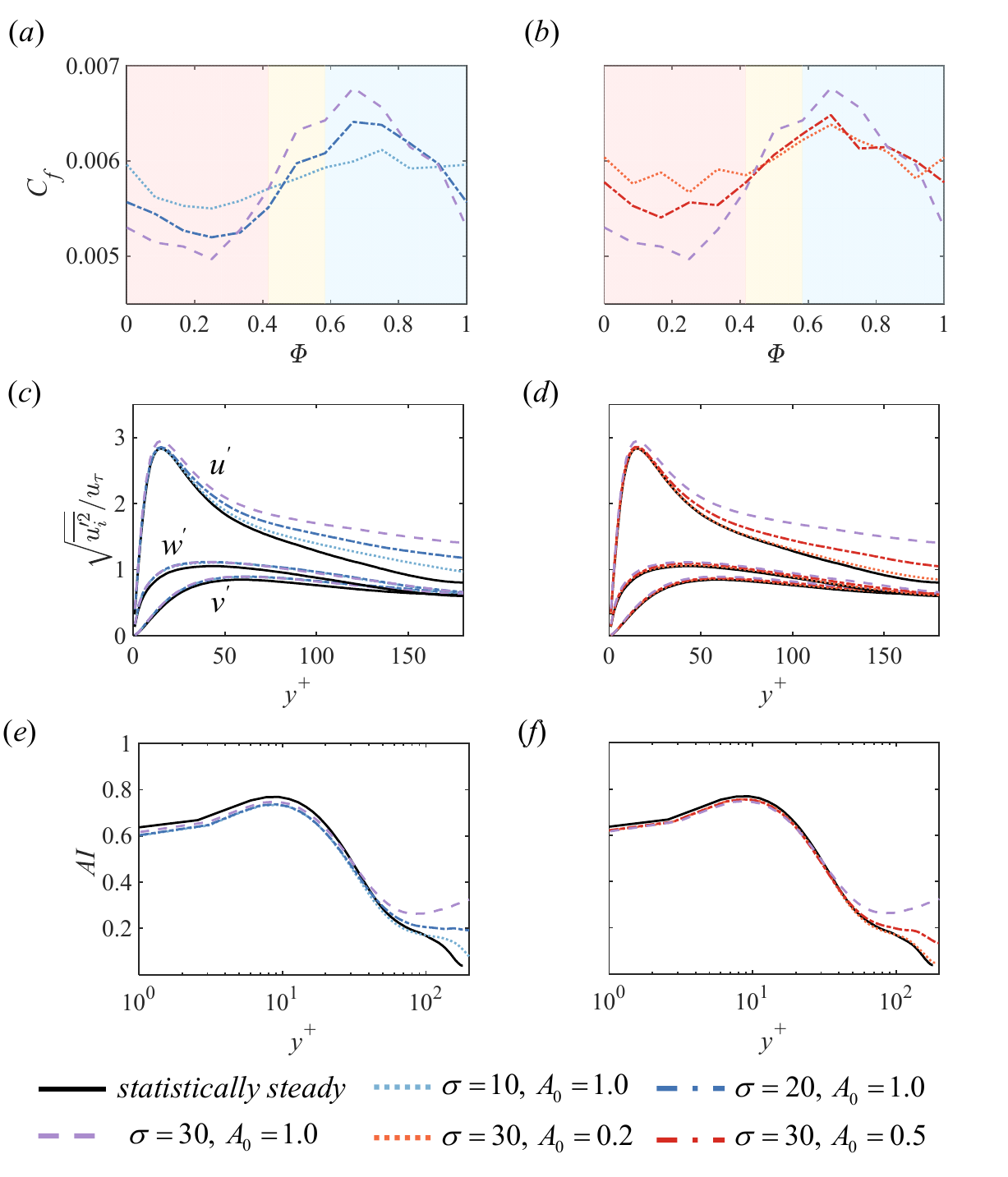}
\caption{\label{fig4} Macroscopic response and modulation of turbulent statistics under transient gust forcing. 
($a$) and ($b$) Phase-averaged skin-friction coefficient $C_f$. 
The background shading qualitatively denotes the acceleration (light red), peak-forcing (light yellow), and deceleration (light blue) phases. 
($c$) and ($d$) Long-time-averaged RMS velocity fluctuations. ($e$) and ($f$) Reynolds-stress anisotropy index (AI). Panels ($a$), ($c$), and ($e$) show the effect of varying gust duration $\sigma$ at $A_0=1.0$; 
panels ($b$), ($d$), and ($f$) show the effect of varying forcing amplitude $A_0$ at $\sigma=30$. The steady baseline is denoted by solid black lines.}
\end{figure}

The non-equilibrium nature of the gust--flow interaction is further examined through the macroscopic response of the wall friction.
Figures~\ref{fig4}($a$) and \ref{fig4}($b$) show the phase-averaged skin-friction coefficient $C_f$ over the gust cycle for different forcing durations $\sigma$ and amplitudes $A_0$.
A clear hysteresis loop is observed, indicating a phase lag between the external forcing and the near-wall turbulent response.
During the acceleration phase ($\Phi<0.42$), $C_f$ remains below the expected quasi-steady response, implying that the near-wall turbulence has not yet adjusted to the rapidly increasing momentum input.
Conversely, during the deceleration phase ($0.58<\Phi<1$), $C_f$ remains elevated, reflecting the pronounced memory effect.
This hysteresis implies that the TKE generated near the gust peak is temporarily stored in large-scale outer-layer structures and subsequently transported toward the near-wall region over a finite relaxation time.

The net structural distortion imprinted on the boundary layer over the gust encounter is further elucidated by the long-time-averaged root-mean-square (RMS) velocity fluctuations, shown in Figs.~\ref{fig4}($c$) and \ref{fig4}($d$).
Presenting these long-time-integrated statistics, rather than only instantaneous peak profiles, highlights the residual kinetic energy and enhanced anisotropy that persist beyond the immediate transient forcing.
As the forcing duration increases at a fixed amplitude [see Fig.~\ref{fig4}($c$)], or as the forcing amplitude increases at a fixed duration [see Fig.~\ref{fig4}($d$)], the streamwise velocity fluctuation $u_{rms}^{\prime}$ intensifies, particularly in the outer region ($y^+>50$).
In contrast, the wall-normal ($v_{rms}^{\prime}$) and spanwise ($w_{rms}^{\prime}$) components exhibit only marginal changes.
This selective enhancement indicates that the gust primarily injects energy into the streamwise component, thereby exacerbating the energetic imbalance among the three velocity components.
To quantify this structural distortion, the Reynolds-stress anisotropy index ($AI$), defined as the second invariant of the anisotropy tensor $II_b$ normalized by its limiting value \cite{lumley1977return,oyewola2004influence,manna2012pulsating}, is evaluated in Figs.~\ref{fig4}($e$) and \ref{fig4}($f$).
The $AI$ profiles further reveal that strong, long-duration gusts ($\sigma=30$, $A_0=1.0$) increase flow anisotropy in the outer layer ($y^+ \sim 100$).
This result suggests that low-frequency, high-amplitude forcing extends the period of energy accumulation, causing large-scale structures to become increasingly strained in the streamwise direction before the energy can cascade to smaller, more isotropic scales.

\begin{figure}
\includegraphics[width=\textwidth]{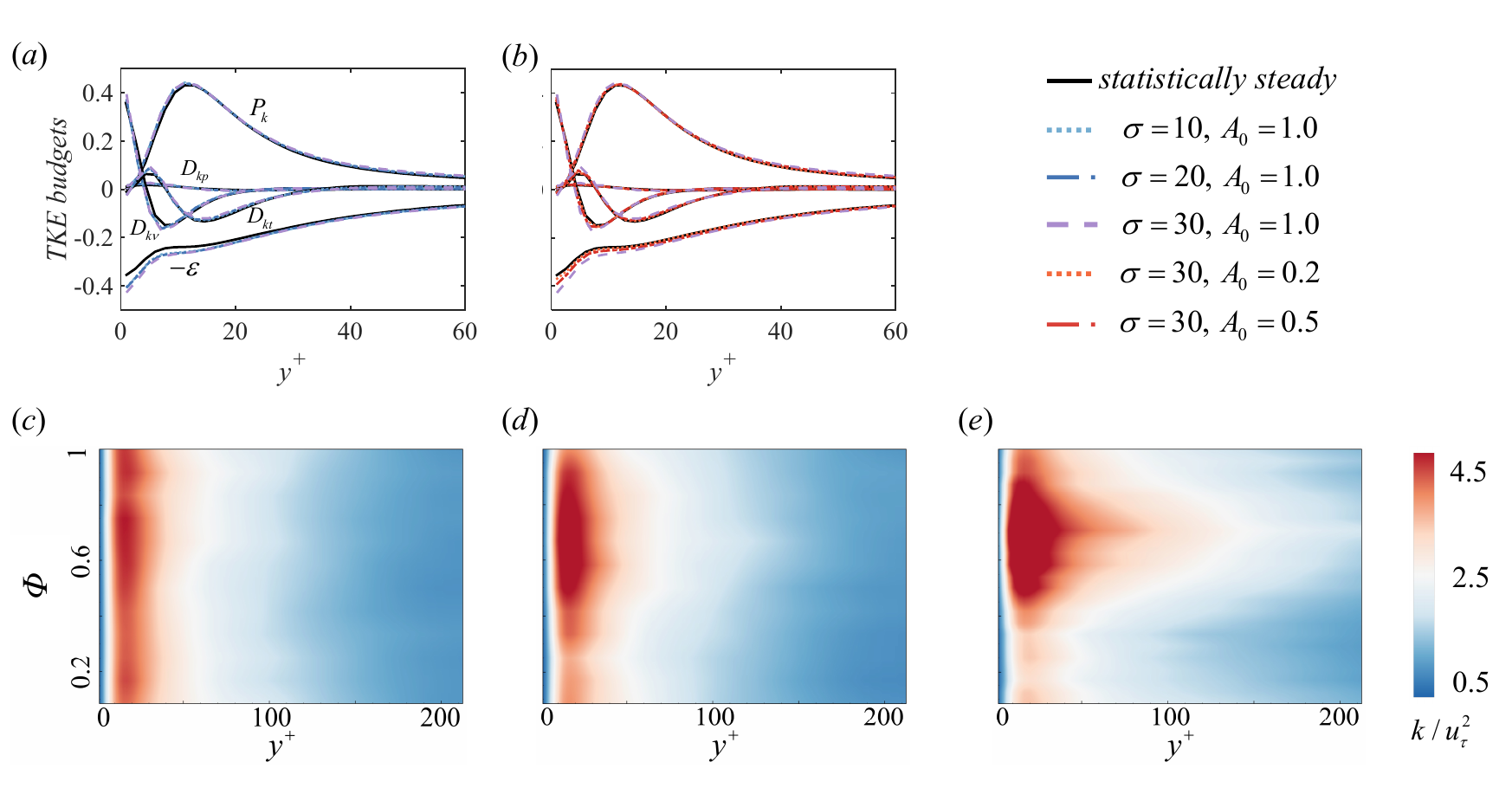}
\caption{\label{fig5} Time-averaged TKE budgets and spatiotemporal evolution of phase-averaged TKE. ($a$) and ($b$) Time-averaged TKE budgets for varying gust durations ($\sigma$) at $A_0=1.0$ 
($a$) and varying amplitudes ($A_0$) at $\sigma=30$ ($b$). ($c$)--($e$) Spatiotemporal contours of normalized phase-averaged TKE for ($c$) short-duration ($\sigma=10$, $A_0=1.0$), 
($d$) long-duration ($\sigma=30$, $A_0=1.0$), and ($e$) extreme ($\sigma=20$, $A_0=5.0$) gusts.}
\end{figure}

The energetic footprint of the gust is further characterized through the TKE budget and the phase-averaged spatiotemporal TKE distribution.
As shown in Figs.~\ref{fig5}($a$) and \ref{fig5}($b$), the time-averaged TKE production and dissipation retain their characteristic near-wall peaks, indicating that the canonical near-wall self-sustaining process is not overridden by the gust.
The time-averaged energy pathways therefore remain dominated by the canonical near-wall cycle.
However, under stronger forcing (e.g., $A_0=1.0$), a slight enhancement of the near-wall dissipation rate ($-\varepsilon$) is observed at $y^+<20$, suggesting a localized gust-induced intensification of small-scale vortical activity.

The most prominent signature of gust modulation is revealed by the phase-averaged TKE contours in the $y^+$--$\Phi$ plane [see Figs.~\ref{fig5}($c$)--($e$)].
For a short-duration gust [$\sigma=10$; see Fig.~\ref{fig5}($c$)], the TKE remains relatively uniform throughout the cycle and acts mainly as a background modulation.
However, as the duration increases to $\sigma=30$ [see Fig.~\ref{fig5}($d$)], the TKE maximum becomes localized within a distinct phase window during the deceleration stage ($0.58<\Phi<1$), accompanied by outward transport of energetic fluid into the buffer layer.
This synchronization suggests that moderate-duration gusts organize the timing of bursting events.
Under extreme gust forcing [$\sigma=20$, $A_0=5.0$; see Fig.~\ref{fig5}($e$)], the turbulent dynamics are substantially reorganized.
The TKE peak shifts toward the late-acceleration and deceleration phases, and the highly energetic region extends beyond the traditional near-wall confinement into the outer layer ($y^+>100$).
This cross-layer energy transfer demonstrates that severe gusts amplify energetic large-scale structures that govern momentum exchange across the boundary layer.

\begin{figure}
\includegraphics[width=\textwidth]{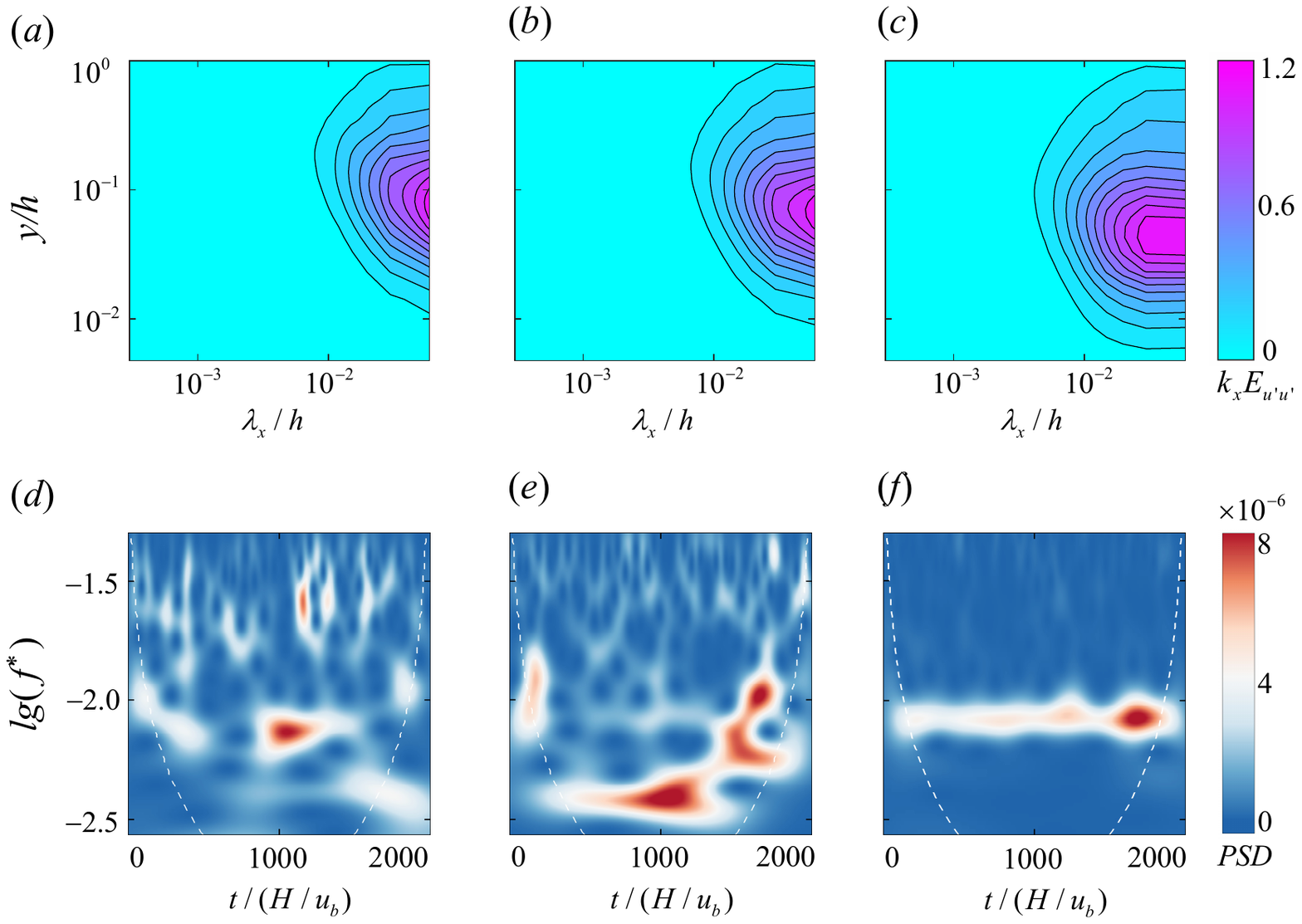}
\caption{\label{fig6} Multiscale energy modulation of streamwise velocity fluctuations. 
($a$)--($c$) Premultiplied 1D energy spectra $k_x E_{u^\prime u^\prime}$ for varying gust amplitudes $A_0$ at fixed $\sigma=20$. 
($d$)--($f$) Local continuous-wavelet power spectral density (PSD) of the streamwise velocity sampled at $y^+ \approx 80$ for the corresponding gust amplitudes. 
White dashed lines indicate the cone of influence (COI). Panels correspond to ($a$) and ($d$) $A_0=0.2$; ($b$) and ($e$) $A_0=1.0$, and ($c$) and ($f$) $A_0=5.0$.}
\end{figure}

To elucidate how the transient gust modulates the distribution of turbulent kinetic energy across spatial and temporal scales, we perform spectral analysis of the velocity fluctuations.
Figures~\ref{fig6}($a$)--\ref{fig6}($c$) present the premultiplied one-dimensional energy spectra of the streamwise velocity, $k_x E_{u^{\prime}u^{\prime}}$ \cite{del2004scaling,smits2011high}, for varying gust amplitudes ($A_0=0.2$, $1.0$, and $5.0$) at a fixed duration ($\sigma=20$).
In the low-amplitude case [see Fig.~\ref{fig6}($a$)], the energy distribution resembles that of a canonical statistically steady channel flow, with a prominent near-wall peak associated with the self-sustaining cycle of typical streak structures \cite{hutchins2007evidence}.
Quantitatively, at $A_0=0.2$, the energy peak of $k_x E_{u^{\prime}u^{\prime}}$ is concentrated at a wavelength of $\lambda_x/h \approx 5\times 10^{-2}$ and a wall-normal height of $y/h \approx 10^{-1}$.
However, as the gust amplitude increases to $A_0=5.0$ [see Fig.~\ref{fig6}($c$)], the energetic peak shifts closer to the wall and broadens toward the long-wavelength, low-wavenumber region.
Specifically, the spatial extent of the peak broadens in the wavelength domain, covering structures within $10^{-2} < \lambda_x/h < 5\times 10^{-2}$, 
and its core location shifts downward to $y/h \approx 4\times 10^{-2}$.
This spectral broadening suggests that severe gust forcing enhances energy accumulation in small-scale near-wall streaks while simultaneously promoting the formation of elongated, energetic streamwise structures. 

Because conventional Fourier spectra cannot resolve the localized and transient nature of the gust, we employ the continuous wavelet transform (CWT) to extract localized time--frequency information from the velocity signals \cite{farge1992wavelet,torrence1998practical,zhou2021wavelet,zhang2025wavelet}. 
The CWT is constructed by translating and dilating a mother wavelet, typically a localized oscillatory function of finite duration, along the time axis. 
This operation generates a family of daughter wavelets at different scales and positions, which are then convolved with the target signal. 
The continuous wavelet transform of the streamwise velocity fluctuation signal $u(t)$ is defined as
\begin{equation}
W(a,b)=\int_{-\infty}^{+\infty}u(t)\psi^*\left(\frac{t-b}{a}\right)\mathrm{d}t.
\end{equation}

Here, $W(a,b)$ denotes the continuous wavelet coefficient. 
The scale parameter $a$ governs the dilation or compression of the wavelet, and thus its effective frequency content, whereas the translation parameter $b$ determines the temporal shift of the wavelet along the signal. 
The term $\psi^*$ denotes the complex conjugate of the mother wavelet function $\psi$. 
To characterize the spatiotemporal distribution of turbulent energy, the local wavelet power spectral density (PSD) is evaluated. 
First, the local energy density of the signal at scale $a$ and time $b$ is obtained from the squared modulus of the wavelet coefficients normalized by the scale parameter, $E(a,b)=|W(a,b)|^2/a$. 
Because the amplitude of the wavelet transform depends on scale, this normalized energy density is further converted into the PSD to remove scale-dependent bias. 
This is achieved by dividing $E(a,b)$ by a factor proportional to the central frequency, yielding
\begin{equation}
\mathrm{PSD}(a,b)=\frac{|W(a,b)|^2}{a}\cdot\frac{2\pi}{N\omega_0},
\end{equation}
where $N$ and $\omega_0$ are parameters associated with the chosen mother wavelet. 
The resulting local wavelet PSD provides a robust representation of the turbulent kinetic energy distribution in both the time and frequency domains. 
However, because the analyzed streamwise velocity signal has a finite temporal duration, edge effects arise near the boundaries of the time series during the convolution process. 
As the scale parameter $a$ increases (corresponding to a lower frequency $f$), the effective temporal support of the mother wavelet broadens proportionally. 
Consequently, the boundary distortions caused by data truncation extend farther toward the center of the signal, forming a distinctive cone-shaped boundary on the time-frequency spectrogram. 
This region is known as the cone of influence (COI) \cite{torrence1998practical,zhang2025wavelet}. 
Within the COI, the calculated wavelet power spectral density is contaminated by edge artifacts, rendering the spectral magnitudes unreliable. 
Therefore, to ensure an evaluation of the multiscale energy distribution, spectral data within the COI must be interpreted with caution.

Figures~\ref{fig6}($d$)--\ref{fig6}($f$) compare the local wavelet PSDs of the streamwise velocity in the logarithmic region ($y^+ \approx 80$) for varying gust amplitudes ($A_0=0.2$, $1.0$, and $5.0$) at a fixed duration of $\sigma=20$.
Under weak forcing [see Fig.~\ref{fig6}($d$)], the turbulent energy remains intermittently distributed over intermediate and high frequencies.
This spectral footprint indicates that the weak gust is insufficient to disrupt the canonical turbulent cycle, and the flow remains dominated by the random bursting events characteristic of statistically steady turbulence.
However, as the forcing amplitude increases to $A_0=1.0$ [see Fig.~\ref{fig6}($e$)], the chaotic broadband fluctuations begin to be suppressed.
Under extreme forcing [$A_0=5.0$; see Fig.~\ref{fig6}($f$)], the high-frequency random fluctuations are suppressed, and the turbulent energy is reorganized into a highly energetic, low-frequency continuous band that spans the entire sampling period.
The emergence of this persistent high-energy structure signifies a qualitative shift in the turbulence dynamics.
A key finding emerges by comparing the frequency of this newly formed energetic band ($f^* \approx 8.5\times10^{-3}$) with the driving frequency of the Gaussian gust ($f_{gust} \approx 8.3\times10^{-3}$).
The close agreement, with a relative deviation of only $2.4\%$, demonstrates that intense periodic forcing dictates the characteristic scale of the dominant coherent structures in the outer layer.
Moreover, the local PSD peak of this reorganized high-energy region reaches $8\times 10^{-6}$, approximately two orders of magnitude higher than the energy of the background high-frequency fluctuations. 
This quantitative evidence shows that intense gusts suppress high-frequency broadband fluctuations and impose their own characteristic timescale, thereby reorganizing turbulent energy into low-frequency structures.
This spectral signature indicates that severe atmospheric disturbances can dictate the dominant temporal and spatial scales of turbulent motions in the boundary layer.

\subsection{\label{sec3.2}Triggering mechanisms and topological structures of extreme events}
\begin{figure}
\includegraphics[width=\textwidth]{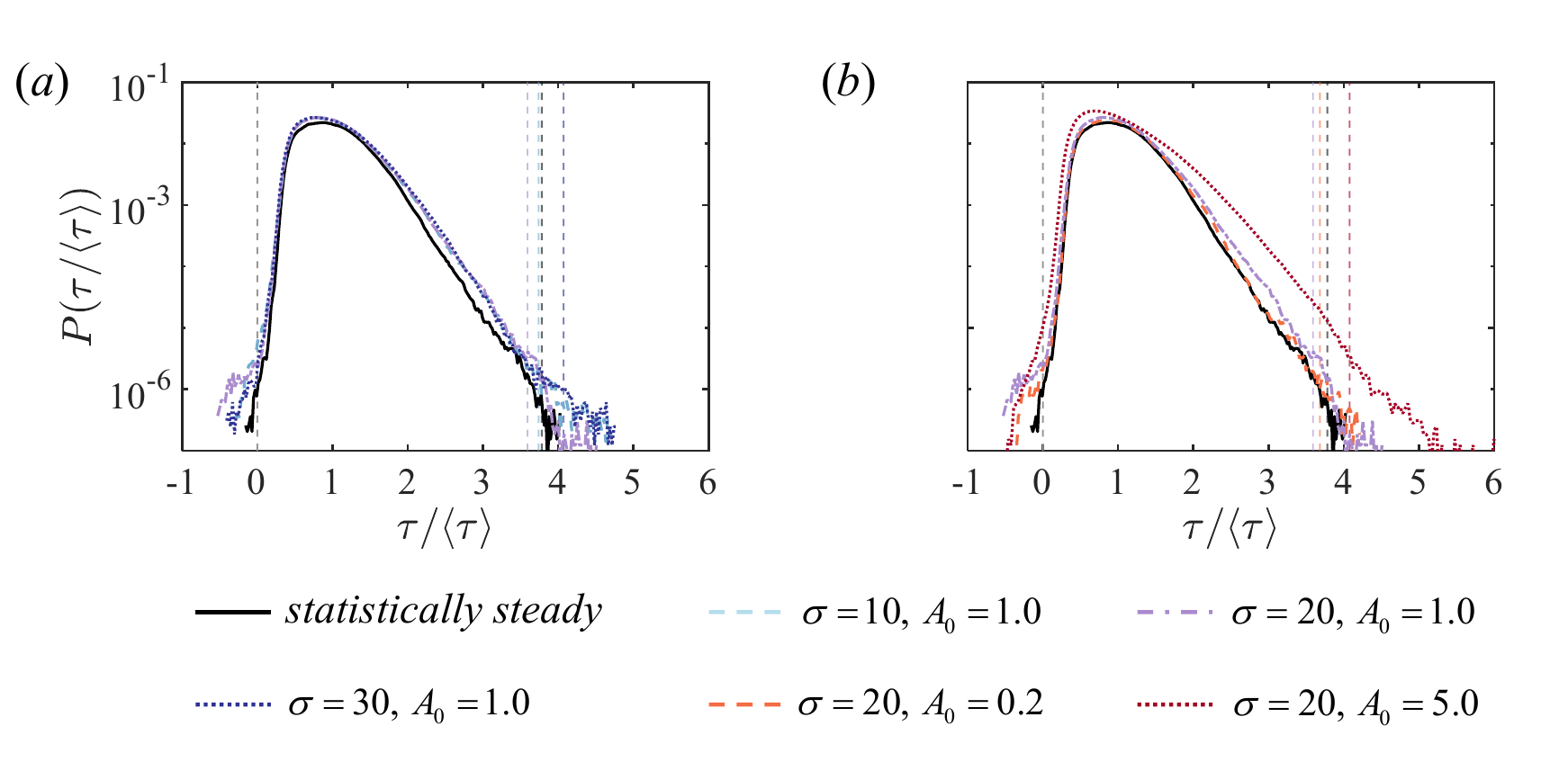}
\caption{\label{fig7} Probability density functions (PDFs) of the normalized instantaneous streamwise wall-shear stress $\tau/\langle\tau\rangle$. 
($a$) Effect of varying gust duration $\sigma$ at $A_0=1.0$. ($b$) Effect of varying amplitude $A_0$ at $\sigma=20$. 
The steady baseline is denoted by solid black lines. Vertical dashed lines indicate the thresholds for extreme positive (EP) events.}
\end{figure}

The flow-structure modulation discussed in Sec.~\ref{sec3.1} suggests that gust forcing alters the probability of extreme near-wall events.
To quantify this intermittency, Fig.~\ref{fig7} presents the probability density functions (PDFs) of the normalized instantaneous streamwise wall-shear stress, $\tau/\langle \tau \rangle$, for both the statistically steady and gust-forced cases.
The heavy tails of the PDFs, which characterize the occurrence of rare events \cite{farazmand2017variational,guerrero2020extreme}, broaden under severe forcing (e.g., $A_0=5.0$).
This broadening indicates an increased probability of both extreme positive (EP) events, corresponding to large positive values of $\tau/\langle \tau \rangle$, and backflow (BF) events, corresponding to $\tau/\langle \tau \rangle < 0$, relative to the canonical statistically steady flow.

\begin{figure}
\includegraphics[width=\textwidth]{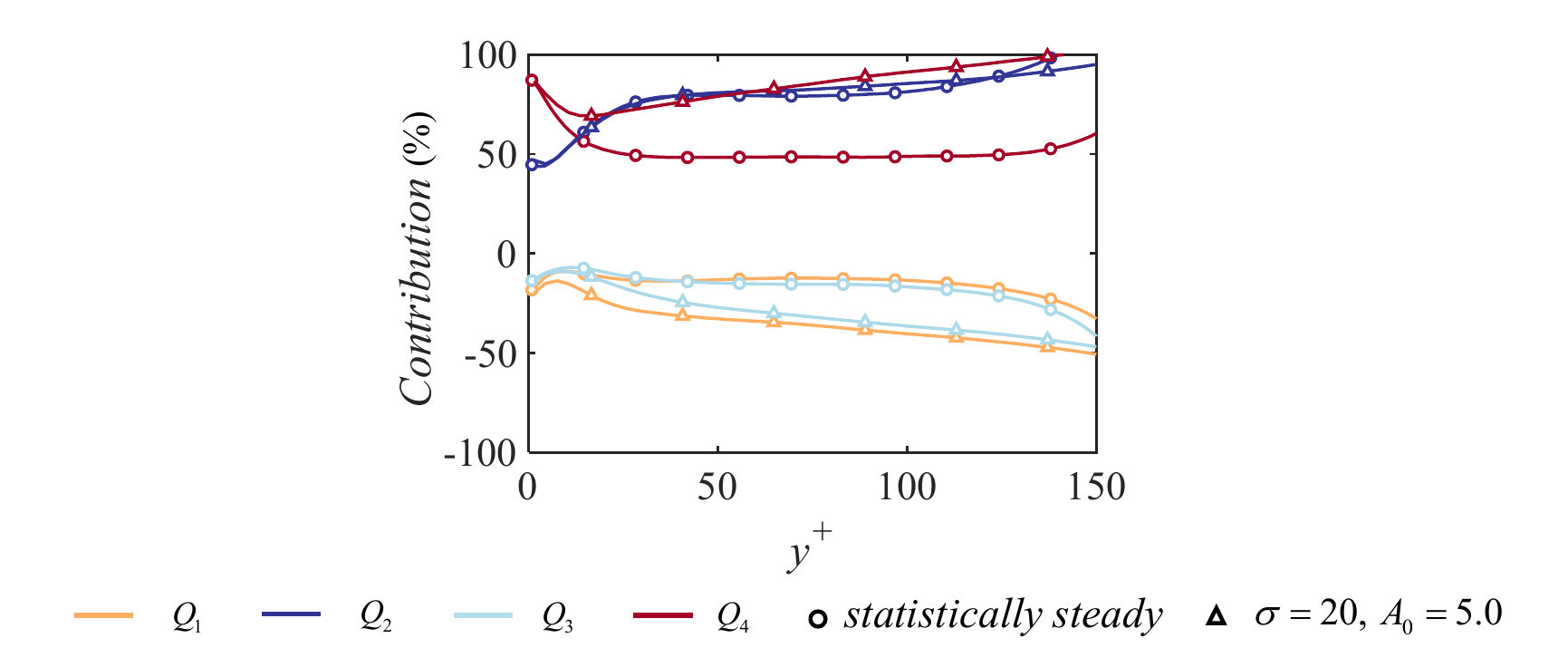}
\caption{\label{fig8} Fractional contributions of the four quadrant events to the mean Reynolds shear stress. 
Circle and triangle symbols denote the steady baseline and the extreme high-amplitude case ($\sigma=20$, $A_0=5.0$), respectively. 
Colors distinguish $Q_1$ (orange), ejection $Q_2$ (dark blue), $Q_3$ (light blue), and sweep $Q_4$ (dark red) events.}
\end{figure}

The dynamic shift underlying these extreme events is elucidated by a quadrant analysis of the Reynolds shear stress, $\langle u^{\prime}v^{\prime}\rangle$ \cite{pan2018extremely}.
Figure~\ref{fig8} compares the fractional contributions of ejection (Q2, $u^{\prime}<0$, $v^{\prime}>0$) and sweep (Q4, $u^{\prime}>0$, $v^{\prime}<0$) events as functions of the wall-normal distance $y^+$.
In statistically steady wall turbulence, it is well established that sweep (Q4) events dominate Reynolds-shear-stress production in the immediate vicinity of the wall ($y^+\lesssim 15$), whereas ejection (Q2) events overtake sweeps and govern turbulent transport throughout the logarithmic and outer regions \cite{wallace1972wall,willmarth1972structure,wallace2016quadrant}.
Under the high-amplitude gust ($A_0=5.0$), however, the contribution of sweep (Q4) events increases substantially.
It surpasses that of ejection (Q2) events, dominates turbulent transport into the outer region ($y^+>20$), and accounts for nearly $80\%$ of the total Reynolds shear stress.
The physical origin of this enhanced sweep (Q4) dominance can be attributed to a gust-induced top-down momentum-transfer mechanism.
When the uniform transient body force is applied, fluid in the outer and bulk regions, which experiences much weaker direct viscous constraint from the wall, accelerates more rapidly than the near-wall fluid.
This differential acceleration generates a large instantaneous mean shear gradient, $\partial U/\partial y$, across the buffer layer.
To restore this macroscopic momentum imbalance, high-momentum fluid from the outer region is driven toward the wall.
Consequently, this non-equilibrium state strongly favors intense inrushes of high-speed fluid (sweeps), suppresses the outward transport of low-speed fluid (ejections), and enhances near-wall momentum exchange.
This shift toward sweep-dominated dynamics serves as the primary physical mechanism triggering EP events.

To uncover the three-dimensional structural origins of these rare, localized phenomena, we employ a two-step conditional-averaging methodology.
Volumetric conditional averaging has recently emerged as a robust tool for extracting coherent structures responsible for extreme near-wall events across various complex flow regimes \cite{lenaers2012rare,cardesa2019structure,guerrero2020extreme,wan2023conditional,wan2025extreme}.
Traditional single-threshold conditional sampling (e.g., averaging all flow fields satisfying $\tau/\langle \tau \rangle < 0$) implicitly assumes statistical symmetry.
However, extreme near-wall events are often driven by asymmetric vortical structures, such as tilted quasi-streamwise vortices or meandering streaks.
Direct ensemble averaging of these randomly oriented asymmetric structures can artificially cancel key kinematic features, thereby smearing their true topological footprint.
To mitigate this issue, the two-step procedure is designed to preserve the spatial coherence of the underlying driving mechanisms.
In the first step, a primary detection box is centered on each identified extreme event, with the local origin defined at $\Delta x^+=\Delta y^+=\Delta z^+=0$.
For backflow (BF) events, the sampling domain spans $\Delta x^+\times\Delta y^+\times\Delta z^+=80\times40\times40$ wall units, ensuring that the immediately overlying turbulent structures are captured;
for extreme positive (EP) events, the domain is adjusted to $60\times30\times30$ wall units to resolve the intense near-wall shear.
In the second step, a secondary detection criterion is applied within the extracted domain to determine the primary rotational orientation of the associated vortical structure.
Specifically, the location of the dominant quasi-streamwise vortex core is identified using the positive second invariant of the velocity-gradient tensor ($Q$-criterion, $Q>0$).
The relative spanwise position of this identified core with respect to the extreme event is then evaluated.
Based on this spatial relationship, the adjacent vortex is classified as either a right-sided ($Q_R$) or left-sided ($Q_L$) quasi-streamwise vortex, depending on whether it lies to the right or left of the event centerline relative to the flow direction.
Before the final ensemble average is performed, the instantaneous flow fields associated with one spatial class (e.g., $Q_R$) are mirrored in the spanwise direction ($z^+$).
This spatial alignment ensures that all conditionally sampled events share a consistent orientation of their driving vortical structures, thereby eliminating destructive interference and yielding a more faithful representation of the extreme-event topology \cite{lenaers2012rare}.

\begin{figure}
\includegraphics[width=\textwidth]{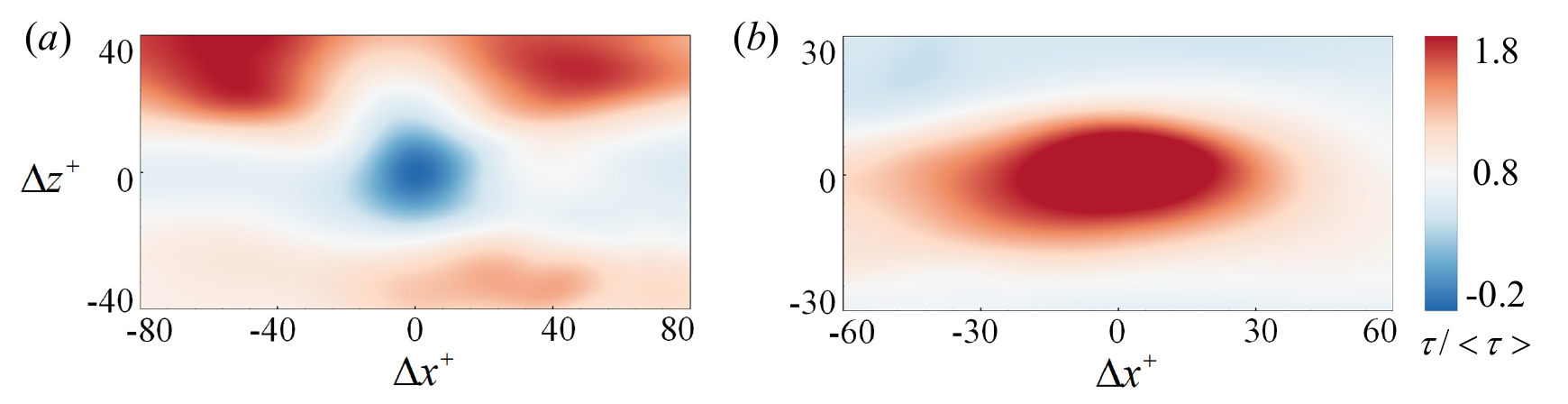}
\caption{\label{fig9} Conditionally averaged contours of the streamwise wall-shear stress 
($\tau/\langle \tau \rangle$) in the $\Delta x^+$--$\Delta z^+$ plane for 
$(a)$ backflow (BF) and $(b)$ extreme positive (EP) events. The fields are extracted under severe gust forcing ($\sigma=30$, $A_0=1.0$), with 
the event centroid located at the origin ($\Delta x^+=\Delta z^+=0$).}
\end{figure}

The parameter space and corresponding statistics of extreme events, including the specific thresholds $\tau_{\text{EP}}^*$ used for event detection, are summarized in Table \ref{tab:table-extreme}. 
Given that BF and EP events are rare phenomena (with occurrence probabilities $P<0.01\%$), conditional averaging in a non-stationary flow requires a large statistical sample.
To ensure statistical convergence and to confirm that the resulting three-dimensional topologies are not artifacts of random fluctuations, the analysis for the severe-gust case ($\sigma=30$, $A_0=1.0$) is accumulated over 25 independent gust cycles.
Within this extended temporal window, a total of 199 independent BF events and 199 independent EP events are identified and analyzed.
This substantial statistical ensemble ensures that random spatiotemporal fluctuations are effectively filtered out.
The effectiveness of this methodology is demonstrated in Fig.~\ref{fig9}, which presents the conditionally averaged footprints of the streamwise wall-shear stress in the wall-parallel $\Delta x^{+}$--$\Delta z^{+}$ plane.
For both BF [Fig.~\ref{fig9}($a$)] and EP [Fig.~\ref{fig9}($b$)] events, the localized, unsmeared core regions confirm that the spatial extent of the underlying structures is well preserved.

\begin{table}[htbp]
\caption{\label{tab:table-extreme} Parameter space and near-wall extreme-event statistics under varying gust conditions, comparing the probabilities of extreme events ($\%EP$ and $\%BF$) and the EP threshold ($\tau_{EP}^\ast$) with the steady baseline.}
\begin{ruledtabular}
\begin{tabular}{ccccc}
$A_0$ & $\sigma$ & $\langle Re_{\tau} \rangle_t$ & $\%EP$ or $\%BF$ & ${\tau}_{EP}^*$\\
\hline
\multicolumn{2}{c}{Statistically steady} & 180.74 & 0.0004\% & 3.7808\\
0.2 & 10 & 187.44 & 0.0017\% & 3.6963\\
0.2 & 20 & 187.06 & 0.0014\% & 3.6821\\
0.2 & 30 & 187.30 & 0.0015\% & 3.7158\\
0.5 & 10 & 198.11 & 0.0017\% & 3.6762\\
0.5 & 20 & 197.94 & 0.0021\% & 3.6476\\
0.5 & 30 & 198.15 & 0.0019\% & 3.7232\\
1.0 & 10 & 214.71 & 0.0022\% & 3.7391\\
1.0 & 20 & 214.35 & 0.0030\% & 3.5910\\
1.0 & 30 & 214.70 & 0.0031\% & 4.0715\\
5.0 & 20 & 317.84 & 0.0041\% & 4.0781\\
\end{tabular}
\end{ruledtabular}
\end{table}

\begin{figure}
\includegraphics[width=\textwidth]{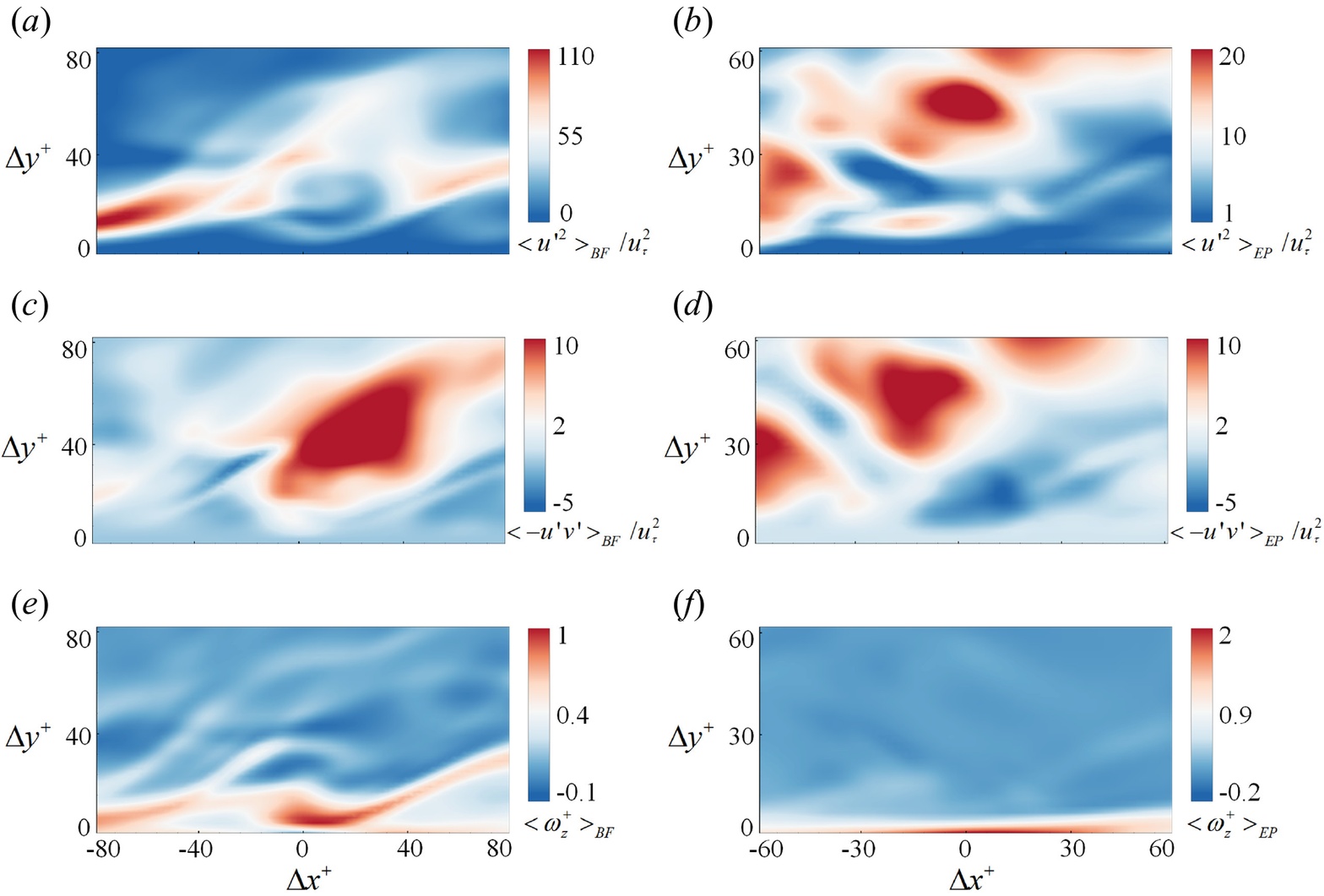}
\caption{\label{fig10} Conditionally averaged topologies of backflow (BF, left column) and extreme positive (EP, right column) events under severe gust forcing ($\sigma=30$, $A_0=1.0$) in the $\Delta x^+$--$\Delta y^+$ plane at $\Delta z^+=0$. 
Normalized conditionally averaged ($a$) and ($b$) streamwise velocity variance, ($c$) and ($d$) Reynolds shear stress, and ($e$) and ($f$) spanwise vorticity for BF and EP events.}
\end{figure}

Figures~\ref{fig10}($a$), \ref{fig10}($c$), and \ref{fig10}($e$) further show the conditionally averaged flow fields centered on the identified BF events in the $\Delta z^+=0$ plane. 
The conditionally averaged streamwise velocity variance [$\langle u^{\prime 2} \rangle_{BF}^{+}$; Fig.~\ref{fig10}($a$)] reveals a region of intensified turbulent fluctuations above the BF centroid ($\Delta y^+>10$). 
Quantitatively, in the region $\Delta y^+ \approx 10 \sim 20$ above the BF event, the conditionally averaged streamwise velocity variance reaches $\langle u^{\prime 2} \rangle_{BF} / u_\tau^2 \approx 110$. 
This value is more than five times the corresponding peak value for EP events, which is only $\langle u^{\prime 2} \rangle_{EP} / u_\tau^2 \approx 20$ [Fig.~\ref{fig10}($b$)]. 
This significant contrast confirms that BF events are associated with intense turbulent fluctuations.
The energetic region extends over a broad wall-normal range, indicating a robust buffer-layer response rather than a localized near-wall anomaly. 
The origin of this enhanced fluctuation activity is further elucidated by the conditionally averaged Reynolds shear stress [Fig.~\ref{fig10}($c$)] and spanwise vorticity $\langle \omega_z^+ \rangle_{BF}$ [Fig.~\ref{fig10}($e$)]. 
Above the BF event ($\Delta y^+>20$), a high-stress zone emerges, while an elongated band of intense positive spanwise vorticity forms near the wall ($\Delta y^+<10$). 
These coupled topological features demonstrate that the BF event is not merely a passive flow reversal; rather, it is associated with a localized adverse pressure-gradient field. 
This gradient, together with the energetic upward ejection of low-speed fluid, promotes the formation of spanwise roll-up vortices, thereby triggering secondary instabilities within the turbulent boundary layer.

The kinematic signature of extreme positive (EP) events contrasts sharply with that of BF events. 
As shown by the conditionally averaged fields in Figs.~\ref{fig10}($b$), \ref{fig10}($d$), and \ref{fig10}($f$), EP events are characterized by high-speed fluid plunging toward the wall. 
Figure~\ref{fig10}($b$) shows the conditionally averaged streamwise velocity variance ($\langle u^{\prime 2} \rangle_{EP}^{+}$). 
A prominent high-intensity fluctuation region descends obliquely from the upper boundary layer ($\Delta y^+ \approx 30\text{--}50$) and impinges on the near-wall region, exhibiting the classic signature of a sweep structure. 
This vigorous downward transport of high-momentum fluid generates a continuous, elongated high-shear channel aligned with the streamwise direction, as evidenced by the Reynolds shear stress distribution in Fig.~\ref{fig10}($d$). 
Although the peak Reynolds shear stress generated by both EP and BF events reaches approximately $\langle -u^{\prime}v^{\prime} \rangle / u_\tau^2 \approx 10$, 
the EP event forms a continuous, wall-inclined high-shear channel that extends streamwise over $\Delta x^+ \approx 40$.
This observation is consistent with the kinematic feature of wall-bounded turbulence in which energetic sweep motions govern extreme positive wall-shear-stress events; 
this mechanism remains robust in both incompressible flows and compressible turbulent boundary layers \cite{wan2023conditional}. 
Consequently, the intense shear at the fluid--wall interface drives substantial local vorticity generation. 
Figure~\ref{fig10}($f$) reveals that the conditionally averaged spanwise vorticity, $\langle \omega_z^+ \rangle_{EP}$, is strongly concentrated in the immediate vicinity of the wall ($\Delta y^+<5$), forming a compressed, intense vortical band. 
Unlike the elevated vortical structures associated with BF events, the extreme vorticity characterizing EP events arises from the localized impact of small-scale, high-speed streaks sweeping onto the wall. 
Quantitatively, EP events are dominated by the sweep of high-speed fluid, which severely compresses intense spanwise vorticity into a thin layer at $\Delta y^+ < 5$, resulting in a conditionally averaged peak vorticity of $\langle \omega_z^+ \rangle_{EP} \approx 2$. 
In contrast, the peak value of the lifted vorticity band induced by BF events is only $\langle \omega_z^+ \rangle_{BF} \approx 1$ [Fig.~\ref{fig10}($e$)]. 
This comparison quantitatively illustrates the dominant role of high-speed sweep streaks in driving momentum exchange during EP events.
The sweeping mechanism dictates the extreme shear gradients and governs the enhanced momentum-exchange pathways under severe gust conditions.

To contextualize these findings within engineering applications, the physical mechanisms elucidated herein inform the underlying dynamics relevant to systems operating in the atmospheric boundary layer, such as low-altitude unmanned aerial vehicles (UAVs). 
Specifically, the pronounced memory effect implies that traditional flight-control algorithms based on quasi-steady assumptions may fail during severe gust encounters. 
Nevertheless, the present study has certain limitations. The spatially uniform temporal surrogate model simplifies the spatial gradients and random amplitude variations of natural gusts. Additionally, owing to the computational cost of DNS, the baseline friction Reynolds number ($Re_\tau \approx 180$) is much lower than that of full-scale flight conditions.

\section{\label{sec4}Conclusions}

In this study, we investigated the response of turbulent shear flow to unsteady, Gaussian-type gust forcing. 
By varying the forcing amplitude $A_0$ and duration $\sigma$, we examined the non-equilibrium modulation of the turbulent kinetic energy (TKE) cascade, the reorganization of coherent structures, and the mechanisms triggering near-wall extreme events. 
The main findings are summarized as follows.

The turbulent response to gust forcing exhibits a pronounced phase lag and hysteresis, indicating a persistent memory effect. 
High-amplitude, long-duration gusts predominantly inject energy into the streamwise velocity component, thereby increasing flow anisotropy in the outer layer $(y^+ \sim 100)$. 
The phase-averaged TKE distribution further reveals that severe gusts drive substantial redistribution of turbulent energy across the boundary layer, shifting the peak turbulent activity from the near-wall region to the outer layer during the deceleration phase.

Wavelet and spectral analyses demonstrate that intense gust forcing dictates the characteristic scale of the dominant turbulent structures. 
Specifically, under extreme gust forcing ($A_0 = 5.0$), the extracted frequency of the newly formed highly energetic band ($f^\ast \approx 8.5 \times 10^{-3}$) deviates by only $2.4\%$ from the gust driving frequency ($f_{\mathrm{gust}} \approx 8.3 \times 10^{-3}$).
The gust suppresses small-scale, high-frequency fluctuations and reorganizes turbulent energy into a highly energetic, low-frequency continuous band governed by the forcing timescale.

Severe gust forcing also increases the probability of near-wall extreme events. 
Quadrant analysis reveals an increased contribution from intense sweep (Q4) motions, which dominate over ejection (Q2) events in the outer region and act as the primary trigger for extreme positive (EP) wall-shear-stress events. 
Using a two-step conditional averaging technique, we elucidate the three-dimensional topologies of backflow (BF) and EP events.
The results demonstrate that BF events are not merely passive flow reversals; instead, they are actively driven by strong, localized adverse pressure gradients and energetic upward ejections $(\Delta y^+>10)$. 
This dynamic coupling induces the formation of intense spanwise roll-up vortices in the buffer layer, thereby triggering secondary turbulent instabilities. 
In contrast, EP events are driven by energetic, high-speed fluid sweeps that plunge obliquely toward the wall. 
Rather than appearing as localized eruptions, these energetic sweeps carve out continuous high-shear channels extending over $\Delta x^+\approx 40$. 
Furthermore, this downward momentum transport compresses intense spanwise vorticity into a concentrated band in the immediate vicinity of the wall ($\Delta y^+<5$). 
This localized compression doubles the peak vorticity magnitude relative to BF events, generating extreme shear gradients and enhancing near-wall momentum exchange.

In summary, this work highlights the impact of transient atmospheric disturbances on the structural equilibrium of wall-bounded turbulence. 
The detailed characterization of non-equilibrium energy transfer and the three-dimensional topology of extreme events provides a theoretical foundation for developing predictive models and robust flow-control strategies for systems operating in complex, unsteady atmospheric environments. 
Despite these insights, the present idealized temporal surrogate model simplifies the random and multiscale characteristics of realistic atmospheric boundary layers. 
Future research will focus on DNS at higher Reynolds numbers and on spatially evolving gust models. 
Moreover, combining the physical mechanisms of cross-scale energy transfer revealed herein with data-driven and machine-learning algorithms offers a promising pathway for developing real-time prediction models and active flow-control technologies to mitigate extreme wall-shear-stress events in flight.

\section*{SUPPLEMENTARY MATERIAL}
See the supplementary material for a video showing the spatiotemporal evolution of coherent structures under the Gaussian gust forcing. 

\begin{acknowledgments}
This work was supported by the National Natural Science Foundation of China (NSFC) through grants nos. 12388101, 12272311, and 12125204; the Young Elite Scientists Sponsorship Program by CAST (Grant No. 2023QNRC001); the 111 project of China (project no. B17037); and the Innovation Capability Support Program of Shaanxi (Program No.2024RS-CXTD-15). The authors acknowledge the Computing Center in Xi’an for providing HPC resources that have contributed to the research results reported within this paper.
\end{acknowledgments}

\section*{Author Declarations}
\subsection*{Conflict of Interest}
The authors have no conflicts to disclose.

\section*{Data Availability}
The data that supports the findings of this study are available from the corresponding author upon reasonable request.

\appendix

\section{Rationale for the gust amplitude $A_0$ and duration $\sigma$}
\label{app:parameter_rationale}

In atmospheric meteorology, gusts are conventionally quantified by the ratio of the instantaneous peak wind speed to the mean wind speed. However, within the framework of canonical channel flow DNS, prescribing an unsteady velocity profile can introduce unphysical spatial distortions. Instead, a time-varying volumetric body force is employed as a temporal surrogate model to investigate transient acceleration effects. Consequently, the parameter $A_0$ represents the dimensionless increment of this driving force. Rather than serving as an arbitrary numerical parameter, $A_0$ acts as a dynamic trigger designed to induce the rapid macroscopic velocity surge characteristic of atmospheric gusts.

\begin{figure}[htbp]
\centering
\includegraphics[width=\textwidth]{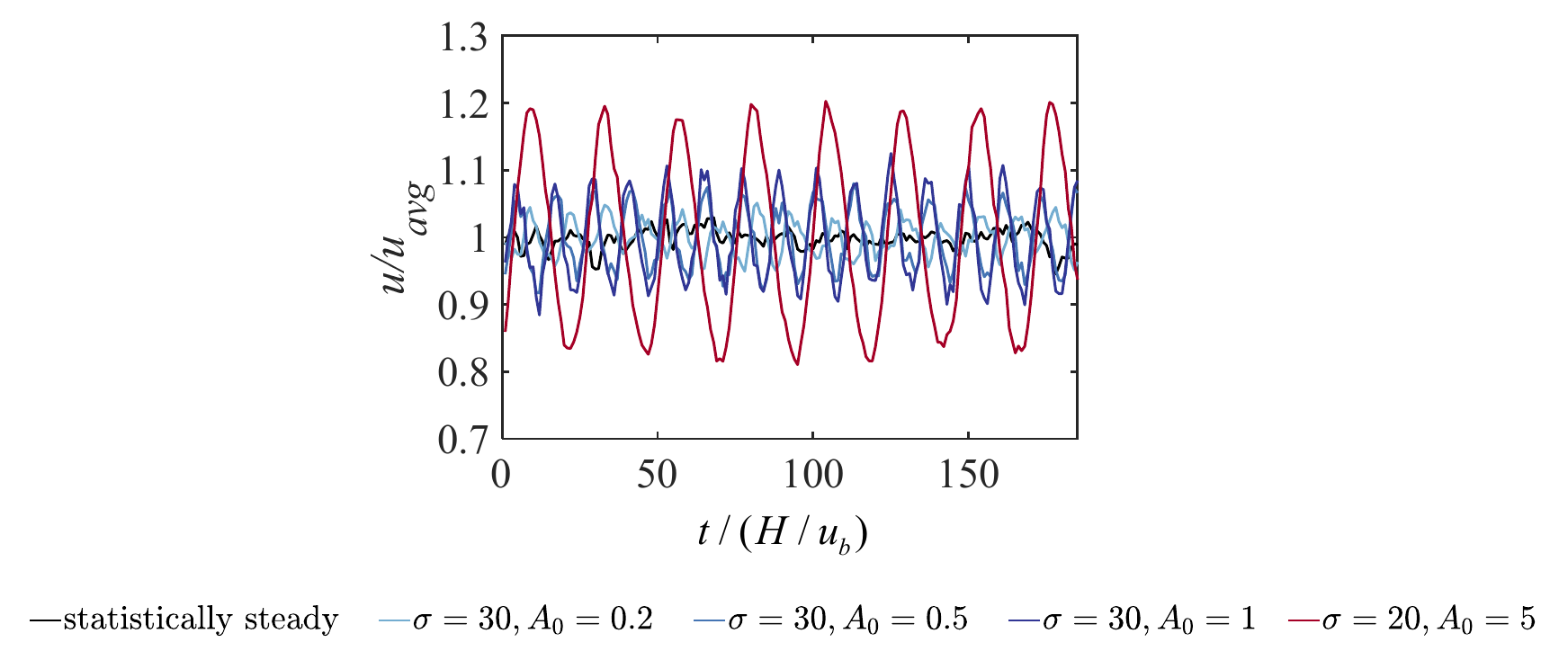} 
\caption{Time evolution of the ratio of instantaneous streamwise velocity to time-averaged velocity ($u/u_{avg}$) at the channel centerline ($y/h=1$). The solid black line denotes the baseline statistically steady flow. The blue lines represent the moderate gust cases with varying amplitudes ($A_0=0.2, 0.5$, and $1.0$) at a fixed duration of $\sigma=30$. The red dashed line represents the extreme gust case with an amplitude of $A_0=5.0$ and a duration of $\sigma=20$.}
\label{fig:A1}
\end{figure}

The operating range of $A_0 \in [0.2, 5.0]$ is constrained by both meteorological relevance and physical turbulence mechanics. Phenomenologically, severe atmospheric gusts are rapid, transient fluctuations from the mean wind speed, with the gust factor (the ratio of peak to mean wind speed) frequently reaching or exceeding 1.5. To evaluate the present forcing model against this kinematic signature, Fig.~\ref{fig:A1} presents the time evolution of the ratio of instantaneous to mean streamwise velocity ($u/u_{avg}$) at the channel centerline ($y/h=1$). For the baseline statistically steady flow ($A_0=0$), this ratio remains stable around 1.0. For moderate forcing amplitudes ($A_0=0.5$ to $1.0$), the velocity ratio reaches values of up to 1.13, exhibiting a typical gust footprint characterized by a sudden velocity surge followed by rapid decay. Under the extreme forcing condition ($A_0=5.0$), the velocity ratio peaks at $\sim$ 1.20. This $20\%$ macroscopic velocity surge is sufficient to disrupt the structural equilibrium of the fully developed turbulent shear flow, driving the system into a non-equilibrium state that captures key features of severe atmospheric disturbances. 

Furthermore, this parameter space respects critical numerical limitations. If $A_0$ is excessively small (e.g., $A_0 \ll 0.2$), the injected momentum is dissipated by the random background turbulent fluctuations and fails to trigger coherent structural reorganization. Conversely, if $A_0$ is excessively large, the resulting extreme velocity surge would induce strong compressibility effects. Within the present incompressible MRT-LBM framework, exceeding the physical limits of the low-Mach-number assumption would violate the governing equations. Thus, $A_0 \in [0.2, 5.0]$ constitutes a physically meaningful and numerically tractable parameter space.

To provide a quantitative rationale for the gust duration parameter $\sigma$, we estimate the timescale separation in a representative atmospheric boundary layer (ABL). Following a standard logarithmic wind profile for a desert surface ($z_0 = 0.01$ m), we consider an atmospheric gust with a mean wind speed corresponding to the lower bound of Beaufort scale level 6 ($U = 10.8$ m/s at $z=10$ m). Integrating the velocity profile over an ABL height of $H=1000$ m gives a friction velocity of $u_\tau \approx 0.63$ m/s and a mean bulk velocity of $u_b \approx 16.4$ m/s. Given the kinematic viscosity of air ($\nu = 1.57 \times 10^{-5}$ m$^2$/s) and a typical meteorological gust duration of $T_{gust} \approx 20$ s, the characteristic timescale measures are $T_{gust}/T_{outer} \approx 0.33$, where the outer eddy turnover time is $T_{outer} = H/u_b$, and $T^+_{gust} \approx 5.0 \times 10^5$ in inner viscous units.

This substantial scale separation highlights a constraint of low-Reynolds-number DNS ($Re_\tau \approx 180$). A significant gap remains between current computational capabilities and the requirements for resolving ABL turbulence under such extreme parameters. Specifically, directly imposing the ABL macroscopic outer timescale ratio ($T_{gust}/T_{outer} \approx 0.33$) in the present DNS would produce an excessively impulsive forcing event spanning fewer than 10 inner time units ($T^+ < 10$), which would artificially disrupt the turbulent structures. Conversely, imposing the inner viscous timescale ($T^+ \sim 10^5$) would render the gust essentially quasi-steady, thereby suppressing the non-equilibrium transient dynamics.


%

\end{document}